\documentclass[journal]{IEEEtran}
\DeclareUnicodeCharacter{0301}{\'}
\DeclareUnicodeCharacter{0300}{\`{}}
\DeclareUnicodeCharacter{221A}{\ensuremath{\sqrt{}}}

\usepackage{todonotes}

\usepackage{geometry}
\usepackage{tabularx}
\usepackage[T1]{fontenc}

\usepackage{amsmath}

\usepackage{lscape}
\usepackage{lipsum} 
\usepackage{rotating} 
\usepackage{afterpage} 
\usepackage{pdflscape} 
\usepackage{float} 

\usepackage{changes}
\definechangesauthor[name={William Lemaire}, color=orange]{WL}

\usepackage{longtable,booktabs,array}
\usepackage{calc} 
\usepackage{etoolbox}
\usepackage{siunitx}
\makeatletter
\patchcmd\longtable{\par}{\if@noskipsec\mbox{}\fi\par}{}{}
\makeatother
\IfFileExists{footnotehyper.sty}{\usepackage{footnotehyper}}{\usepackage{footnote}}
\makesavenoteenv{longtable}

\usepackage{graphicx}

\makeatletter
\def\maxwidth{\ifdim\Gin@nat@width>\linewidth\linewidth\else\Gin@nat@width\fi}
\def\maxheight{\ifdim\Gin@nat@height>\textheight\textheight\else\Gin@nat@height\fi}
\makeatother
\setkeys{Gin}{width=\maxwidth,height=\maxheight,keepaspectratio}
\makeatletter
\def\fps@figure{htbp}
\makeatother
\ifCLASSINFOpdf
\else
\fi
\begin{document}
%
\title{Design and Implementation of a Low-Power Low-Noise Biopotential Amplifier in 28 nm CMOS Technology with a Compact Die-Area of 2500~\si{\micro\meter\squared}}
%
%
%

\author{Esmaeil~Ranjbar~Koleibi, 
    Konin~Koua,
    William~Lemaire,
    Maher~Benhouria,
    Marwan~Besrour,
    Takwa~Omrani,
    Jérémy~Ménard,
    Louis-Philippe~Gauthier,
    Montassar~Dridi,
    Mahziar~Serri~Mazandarani,
    Benoit~Gosselin~\IEEEmembership{Member,~IEEE,}
    Sébastien~Roy~\IEEEmembership{Member,~IEEE,}
    and~Réjean~Fontaine~\IEEEmembership{Senior Member,~IEEE}
}
%
%

\markboth{Journal of \LaTeX\ Class Files,~Vol.~XX, No.~X, August~202X}%
{Shell \MakeLowercase{\textit{et al.}}: Bare Demo of IEEEtran.cls for IEEE Journals}
%



\maketitle

\begin{abstract}
This paper presents a compact low-power, low-noise bioamplifier for multi-channel electrode arrays, aimed at recording action potentials. The design we put forth attains a notable decrease in both size and power consumption. This is achieved by incorporating an active lowpass filter that doesn't rely on bulky DC-blocking capacitors, and by utilizing the TSMC 28 nm HPC CMOS technology. This paper presents extensive simulation results of noise and results from measured performance. With a mid-band gain of 58 dB, a -~3~dB bandwidth of 7 kHz (from 150 Hz to 7.1 kHz), and an input-referred noise of 15.8~\si{\micro\volt_{rms}} corresponding to a NEF of 12. The implemented design achieves a favourable trade-off between noise, area, and power consumption, surpassing previous findings in terms of size and power. The amplifier occupies the smallest area of 2500~\si{\micro\meter\squared} and consumes only 3.4~\si{\micro\watt} from a 1.2 V power supply corresponding to a power efficiency factor of 175  and an area efficiency factor of 0.43, respectively. 

\end{abstract}


%
\IEEEpeerreviewmaketitle

\section{Introduction}
%
%
%
%

\begin{figure}[]
\centering
  \includegraphics[width=0.4\textwidth,height=0.4\textheight,keepaspectratio]{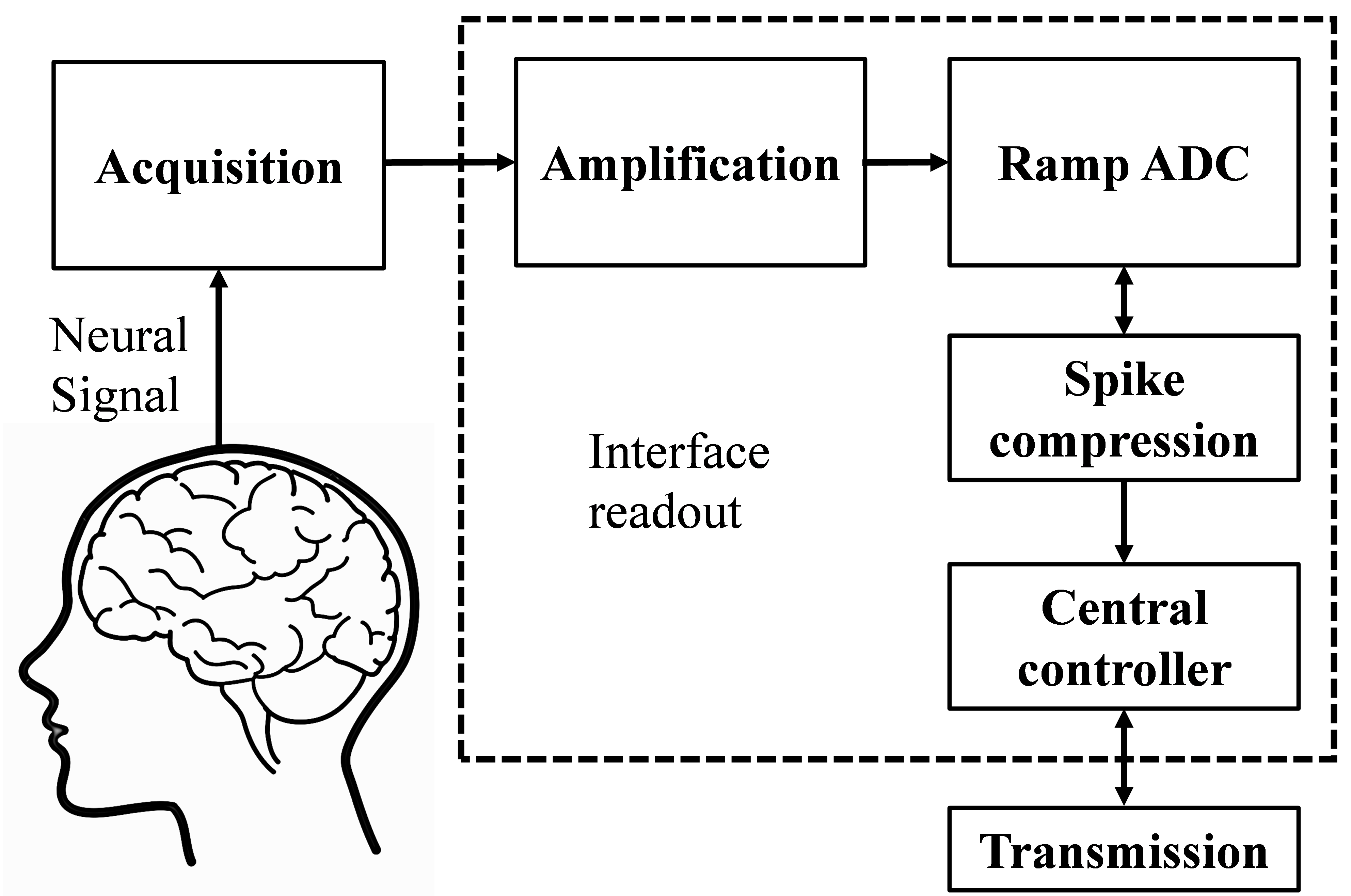}
 
  \caption{Block diagram of an implanted neural recording system comprising of acquisition, interface readout, and transmission stages~\cite{lemaire2022preliminary}.}
    \label{NRS}
\end{figure}

A multielectrode neural recording is nowadays crucial for neuroscience researchers and clinicians to provide significant insights into the physiology of the human body~\cite{raducanu2016time,buzsaki2015tools}. This technology provides means to investigate the operation of the nervous system, comprehend the underlying mechanisms of diverse neurophysiological behaviours, and address neurological disorders like Parkinson's disease, Alzheimer's disease, and epilepsy~\cite{ballini20141024,zou2013100,limousin1998electrical}. Fig. \ref{NRS} illustrates a block diagram of the different implanted parts of a neural recording system comprising blocks such as acquisition, interface readout, and transmission.
   
For the prolonged treatment of chronic diseases, a neural amplifier must fulfil several essential requirements. These encompass 1) a substantial input impedance, 2) minimal power usage to prevent tissue harm and optimize battery life~\cite{shahrokhi2010128,mohseni2004fully,gosselin2007low,chandrakumar2017high}, 3) compact area for system miniaturization, 4) low-level of input-referred noise for accurate spike detection amidst background noise, 5) substantial common-mode rejection ratio, 6) elimination of DC voltage~\cite{jomehei2019low}. Furthermore, the need for electrode arrays with increased density is on the rise. To fulfil these requirements, the interface necessitates low-noise amplifiers (LNAs), filters, and data converters, along with processing circuitry, such as neural spike detectors for in vivo data reduction~\cite{lemaire2022preliminary,8325072}.

This paper discusses the development and assessment of a low-power integrated low-noise biopotential amplifier. This bioamplifier plays a key role in a 49-channel neural recording Application-Specific Integrated Circuit (ASIC) crafted using TSMC CMOS 28 nm technology. Notably, the ASIC integrates an active mechanism for suppressing low-frequency signals.

The bioamplifier demonstrates a bandpass frequency response and bypasses the need for passive AC coupling input networks. An important achievement in this design is the attainment of one of the most condensed footprints noted in literature. This accomplishment is complemented by successfully maintaining a harmonious blend of noise attributes and power consumption.

The subsequent sections of this paper are organized as follows. Section 2 provides a detailed description of the proposed architecture, beginning with an explanation of the system requirements and followed by a demonstration of the proposed design. In Section 3, the circuit details are elaborated upon. For the simulation and measurement of the LNA, the materials used are outlined in Section 4, while the methods developed for simulating and measuring the results are presented in Section 5. Section 6 presents the design results, and a comparative analysis with the state-of-the-art is provided in Section 7. Finally, Section 8 offers the concluding remarks for this paper.

\hfill 
 
\hfill

\section{ARCHITECTURE OF THE BIOPOTENTIAL AMPLIFIER}
\subsection{ System requirements }

The biopotential amplifier is subject to requirements such as signal processing, power consumption silicon area optimisation and DC-offset elimination. These constraints are further elaborated.

Signal Processing: The amplifier must proficiently process action potentials (AP) – low-frequency bioelectric signals primarily concentrated between 300~Hz and 10~kHz~\cite{wu2005multi}. While the peak amplitude of APs generally remains within a few tens of micro-volts, in abnormal scenarios involving superimposed activity from multiple neurons, it could surge up to several millivolts. Given the susceptibility of this frequency range to high levels of noise spectral density in MOSFET devices, prioritizing noise reduction is of utmost importance~\cite{wu2005multi,motchenbacher1993low,zou2013100}.

Power Consumption: The amplifier must strictly adhere to a power consumption limit of 0.8~\si{\milli\watt\per\milli\meter\squared} in chronically implanted devices, ensuring that the temperature rise stays below 1~\si{\celsius}. This constraint is critical to prevent necrosis in muscle tissue and extend the device's battery life~\cite{seese1998characterization}.

Silicon Area Optimization: The amplifier should optimize the silicon area to accommodate the increasing number of electrodes in a Multichannel Electrode Array (MEA). As the electrode count grows, the recording interface pixel area must decrease to maintain a compact recording system size. This necessity arises from the demand for dedicated electronic circuitry for each electrode, involving signal amplification, filtering, and multiplexing.

DC-Offset Elimination: The amplifier must eliminate DC-offset voltages as high as 2 V that may emerge at the recording system's input due to the electrochemical nature of the electrode-tissue interface~\cite{ferris1978introduction}. Eliminating this offset is crucial to avoid amplifier circuit saturation and ensure precise signal processing. A comprehensive analysis of this requirement is provided in the subsequent subsection.

\subsection{Low-Frequency Suppression}

Various biopotential amplifier concepts enable low-frequency suppression through different methods. First, there are capacitor feedback networks~\cite{lee201064,yazicioglu2008200,chen2013fully,shahrokhi2010128}. Second, RC high-pass filtering involves leveraging both the electrode-electrolyte capacitance and a substantial resistor placed between the bioamplifier input and the ground. Third, AC-coupling is achieved by connecting large capacitors in series with the input electrode~\cite{nagel2000biopotential}. These three approaches, characterized by their simple architectures, offer excellent noise performance with minimal power consumption.

Additionally, there are alternative methods. Fourth, active AC coupling employs a digital feedback~loop~\cite{kassiri2016battery,muller20110}. Fifth, active DC coupling utilizes closed-loop regulation of the DC level~\cite{spinelli2004novel,gosselin2007low}. These methods are known for their compact designs.

Finally, the sixth method involves a differential difference amplifier configuration~\cite{rodriguez2012low}, which can enhance common-mode parameters and address supply noises.

Nevertheless, each concept faces limitations that impede the creation of an ideal biopotential amplifier.

- AC-coupling is susceptible to charging effects, occupies a significant die area, and may necessitate off-chip passive components to achieve high-value capacity.

- RC filtering demands large integrated resistors and biasing circuits.

- Impedance mismatches in passive components can decrease Common-Mode Rejection Ratio.

- The cut-off frequency of bioamplifiers, which relies on electrode characteristics, may vary due to electrodeposition and electrical parameters.

- Additional active networks, such as active feedbacks, can lead to elevated power consumption and inadequate noise performance in bio-amplifiers.

\begin{figure}[]
  \includegraphics[width=0.45\textwidth,height=0.45\textheight,keepaspectratio]{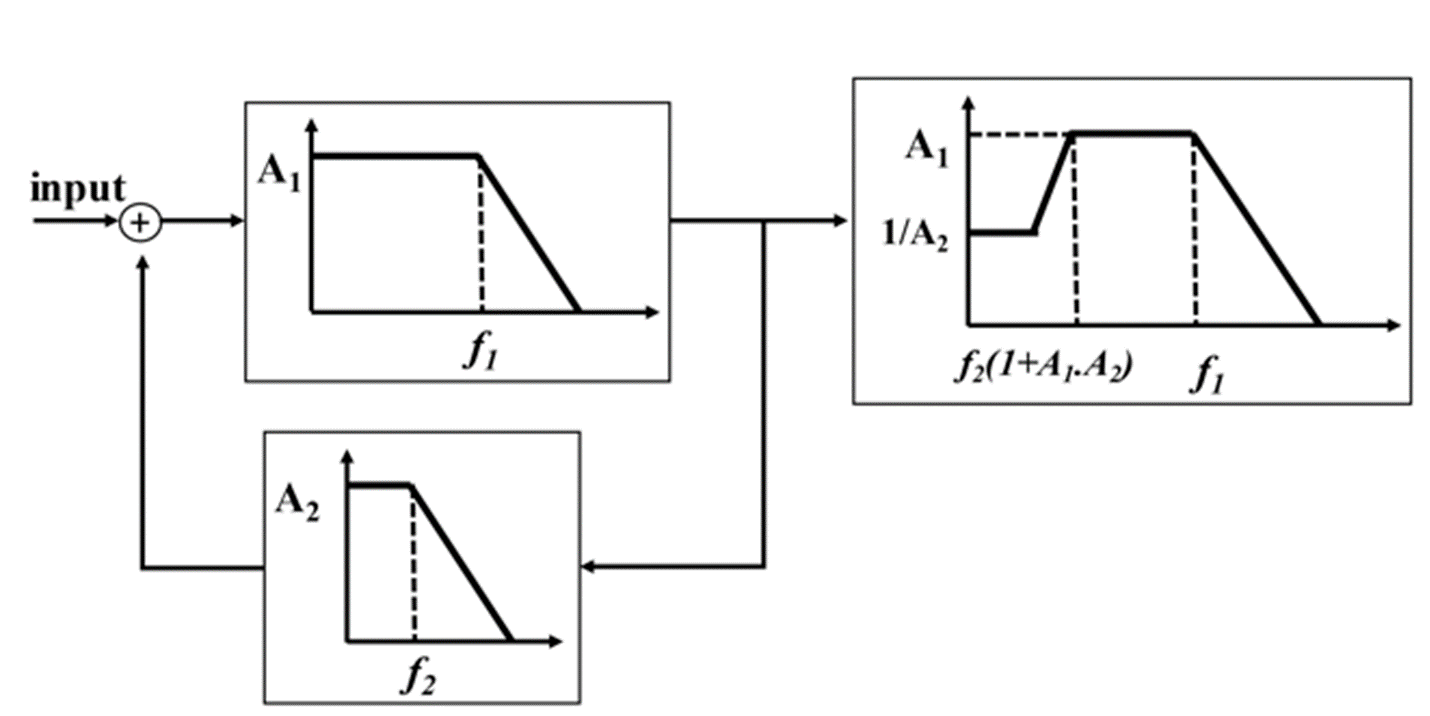}
  \centering
  \caption{Frequency response of high-Pass filter implementation using low-pass filter in feedback.}
    \label{LNA_freq}
\end{figure}

In this paper, we present a DC-coupled active amplifier that employs a closed-loop mechanism to reduce chip area and optimize power consumption, all while maintaining a satisfactory noise level. This approach ensures enhanced input impedance and minimal signal attenuation in contrast to the capacitive feedback network. By incorporating low-pass filtering within the feedback configuration, the method efficiently addresses the issue of DC offset voltage, resulting in substantial DC attenuation within the system transfer function~\cite{chandrakumar2017high,spinelli2004novel}. The low-pass filter monitors the DC current at the amplifier output and performs a subtraction operation on the input, thereby establishing a distinct high-pass characteristic at the system level (Fig.~\ref{LNA_freq}). Although the implementation of this approach necessitates additional active circuitry for high-pass filtering within the feedback loop, leading to a rise in power consumption, its merits include superior referenced noise at the input~\cite{bafar2013wireless}, enhanced DC rejection, and a compact design footprint, making it an extremely promising solution for biopotential amplifiers~\cite{gosselin2007low}.

\section{BIOAMPLIFIER CIRCUIT DESIGN}

\subsection{Analyzing the Frequency Response and Circuit Characteristics of the Biopotential Amplifier}

\begin{figure}[]
  \includegraphics[width=0.4\textwidth,height=0.4\textheight,keepaspectratio]{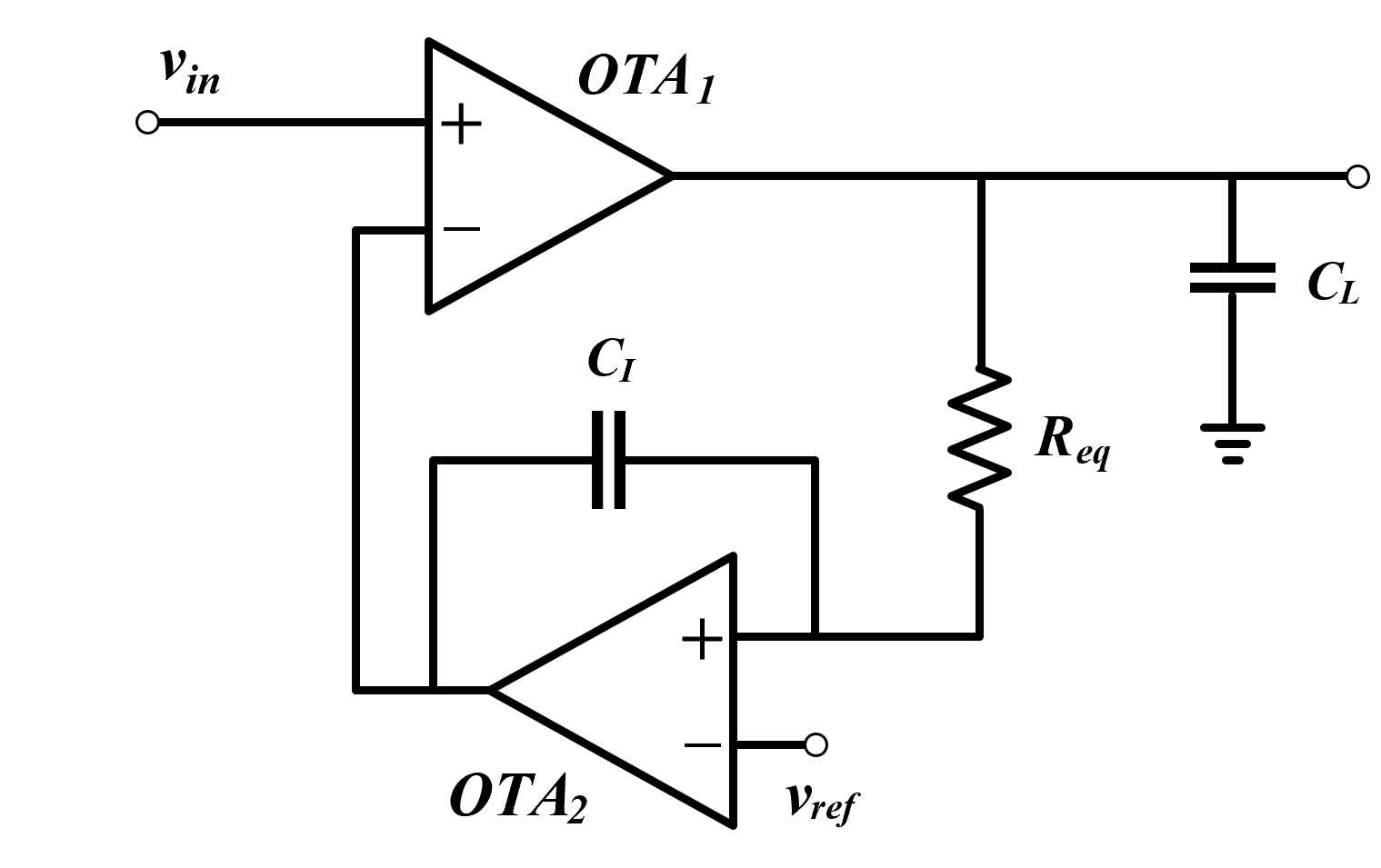}
  \centering
  \caption{Proposed LNA systemic schema with high-Pass filter implemented through low-pass filter feedback.}
    \label{LNA_sys}
\end{figure}

The proposed biopotential amplifier's system diagram is depicted in Fig.~\ref{LNA_sys}, consisting of two single-ended operational transconductance amplifiers (OTA). The amplifier's output typically encounters a load capacitance C\textsubscript{L} of around 150~fF. The feed-forward amplifier OTA\textsubscript{1} establishes the LNA's low-frequency gain and low-pass cut-off frequency (f\textsubscript{1}), which consequently determines the pass-band gain.

The active Miller integrator in the feedback network comprises OTA\textsubscript{2}, a capacitor, and a resistor with a high value. The RC time constant (\si{\tau}) plays a role in controlling the \(-3~dB\) high-pass cut-off frequency of the bio-amplifier \((\tau=R_{eq} .C_I)\). To achieve the desired high-pass cut-off frequency, both R\textsubscript{eq} and C\textsubscript{I} need to have high values~\cite{koleibi2022low}.

\begin{figure}[]
  \includegraphics[width=0.4\textwidth,height=0.4\textheight,keepaspectratio]{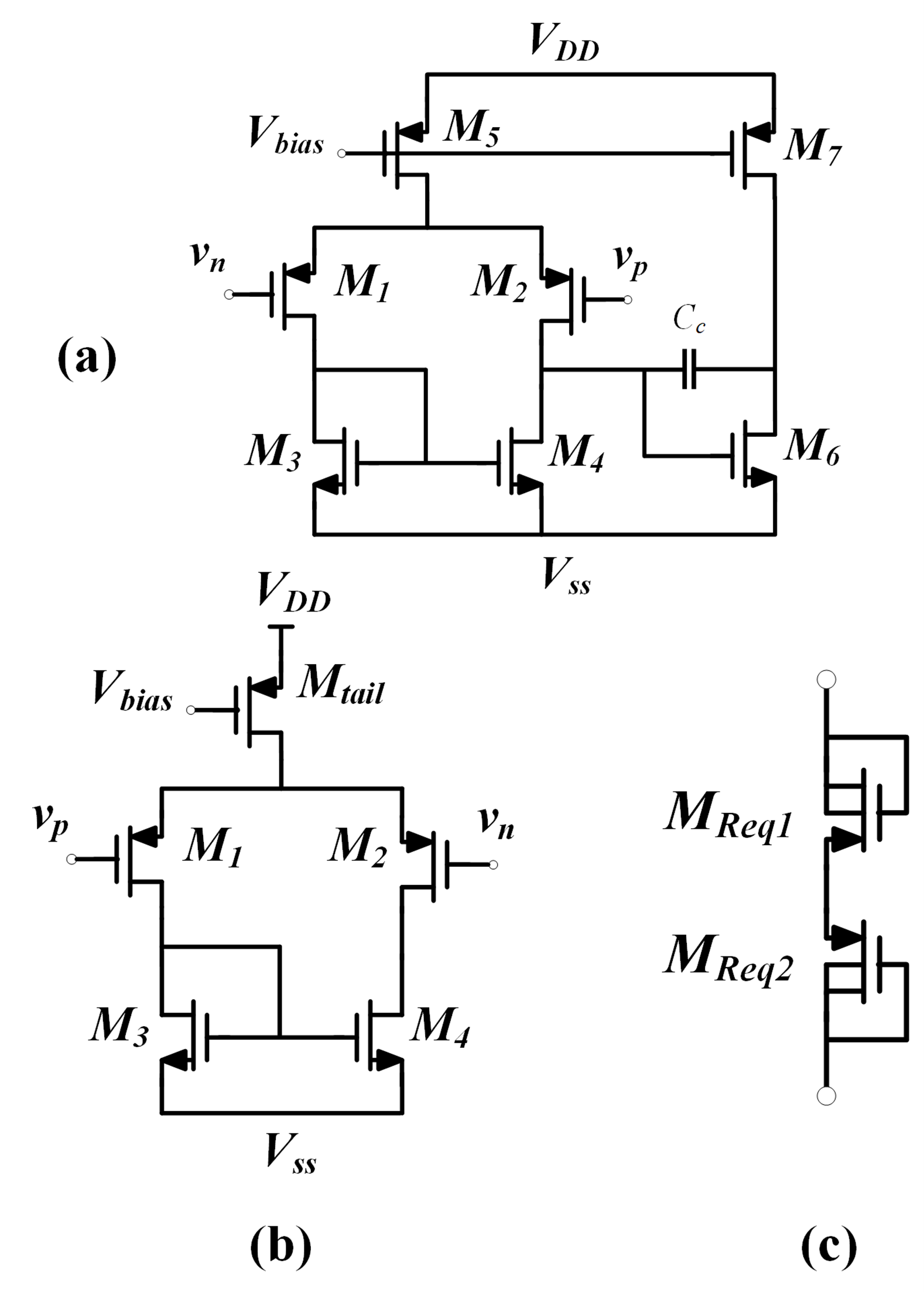}
  \centering
  \caption{Circuit diagrams of (a) \(OTA_1\), (b) \(OTA_2\), (c) \(R_{eq}\).}
    \label{LNA_Circuit}
\end{figure}

For R\textsubscript{eq}, it is built using a non-tunable Quasi-Infinite Resistor (QIR) based on the Two Series Connected Outwardly with a Connected Gate MOS (TSOCGM) structure~\cite{zhao2009low}, as seen in Fig.~\ref{LNA_Circuit} (c). The QIR design guarantees a stronger resistance throughout the useful voltage range, thanks to its symmetric architecture. This characteristic makes the pseudo resistor's architecture less susceptible to nonlinear effects on the LNA's performance. A comparison of the current-voltage characteristics of the proposed QIR with the conventional design can be found in reference ~\cite{sharma2021mos}. Furthermore, in this study employing 28 nm CMOS technology, transistors with a relatively high threshold voltage are utilized to reduce leakage, resolving concerns related to current leakage in QIR implementation. This results in a steady output voltage with minimum variations, free from any voltage fluctuations associated with the QIR.

\subsection{Input-Referred Noise}

Eq.~\ref{eq_noise} represents the total power of input-referred noise of the proposed LNA in \emph{V\textsubscript{rms}}~\cite{motchenbacher1993low,aziz2009256} which consists root mean square (RMS) values of thermal noise component (\(\overline{v_{th}}\)) and flicker noise component (\(\overline{v_{f}}\)) components for OTA\textsubscript{1} and OTA\textsubscript{2}.

The proposed LNA's total power of input-referred noise is described by Eq.~\ref{eq_noise}, measured in \emph{V\textsubscript{rms}}~\cite{motchenbacher1993low,aziz2009256}. It comprises the RMS values of the thermal noise component and flicker noise component for OTA\textsubscript{1} and OTA\textsubscript{2}. Meanwhile, this equation shows that in the suggested circuit, connecting the output of OTA2 to the input of the LNA means that the input-referred noise of the LNA is influenced by both the input-referred noise of OTA\textsubscript{1} and the output noise of OTA\textsubscript{2}.

\begin{equation} \label{eq_noise}
\begin{aligned}
v_{\text{in,total,rms}}^2 &= v_{\text{in1,th,rms}}^2 + v_{\text{in1,f,rms}}^2 \\
&\quad + v_{\text{out2,th,rms}}^2 + v_{\text{out2,f,rms}}^2
\end{aligned}
\end{equation}

The examination of differential amplifiers' noise reveals that the primary noise originates from the input pair of the op-amp and current-mirror pair, neglecting the short-channel effect. Eq.~\ref{eq_Noise_Fl&ther} presents a comprehensive breakdown of the flicker and thermal noise contributions for one OTA. In order to minimize the input-referred noise of each OTA, it is essential to consider certain general factors. Firstly, biasing the input pair transistors in the subthreshold region proves effective in reducing thermal noise, maximizing the \emph{g\textsubscript{m}/I\textsubscript{D}} parameter. Secondly, incorporating large PMOS transistors (\emph{M\textsubscript{1}, M\textsubscript{2}}) with significant \emph{W} and \emph{L} values in the input pair aids in minimizing flicker noise in both OTAs. These low-noise requirements result in considerable allocations of resources in terms of silicon surface area and power consumption for LNAs.

Adopting very short-channel technology offers advantages in terms of area and power reduction. As transconductance directly correlates with transistor current, employing low-voltage very short-channel transistors presents a viable option for reducing input-referred noise simply by increasing the current. This eliminates the need for extra-large transistors (\emph{WL\textsubscript{1,2}, WL\textsubscript{3,4}}) in the amplifier's input stages.

\begin{multline}\label{eq_Noise_Fl&ther}
\overline{V^2_{n,in}} = \underbrace{8kT \left(\frac{2}{3g_{m1}} + \frac{2g_{m3}}{3g_{m1}^2}\right)}_\text{Thermal Noise} \\
+ \underbrace{\frac{2K_P}{C_{ox}(WL)_1 f} + \frac{2K_N}{C_{ox}(WL)_3 f} \frac{g_{m3}^2}{g_{m1}^2}}_\text{Flicker Noise}
\end{multline}

\section{Hardware Used in Biopotential Amplifier Measurement}

\subsection{Simulators}

Cadence Spectre APS simulator was used for the simulation of integrated circuit (IC) performance, covering DC, AC, transient, and noise parameters. Cadence Virtuoso was employed for schematic and layout design.

\subsection{Test Printed Circuit Board (PCB)}

\begin{figure}[]
  \includegraphics[width=0.4\textwidth,height=0.4\textheight,keepaspectratio]{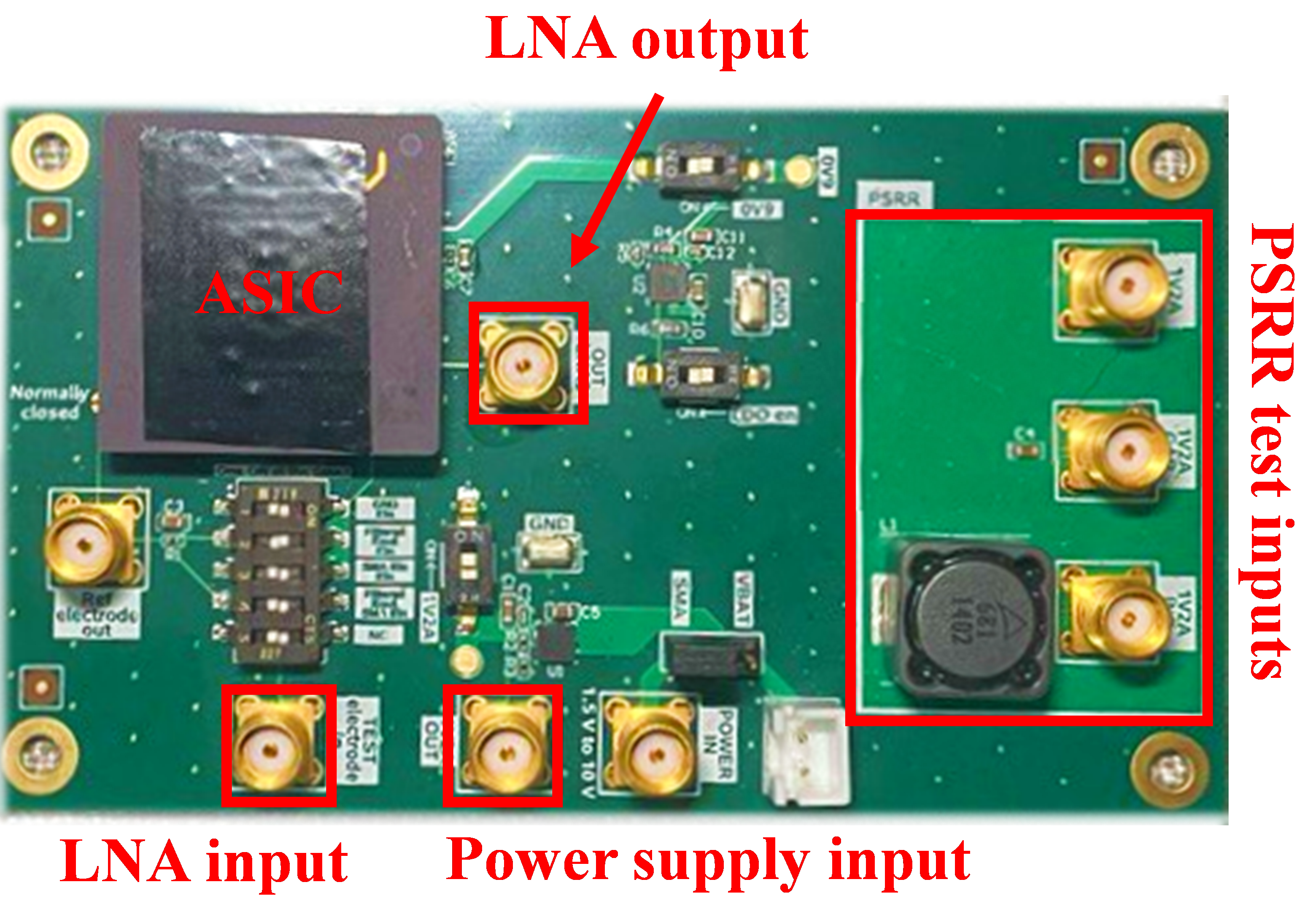}
  \centering
  \caption{Test PCB of the LNA.}
    \label{PCB}
\end{figure}

The PCB depicted in Fig. \ref{PCB} was purposefully designed to evaluate the performance of the LNA while effectively mitigating the impact of noise originating from the digital circuitry of the ASIC. It is a four-layered board that can be powered either by a battery or by an external voltage via an LDO regulator. The board incorporates multiple SMA connectors to optimize measurement accuracy and minimize noise interference.

\subsection{Cables and Connectors}
\subsubsection{SMA Connector Receptacles} The PCB integrates 50~\si{\Omega} SMA connector receptacles, beneficial for low-noise applications due to their superior shielding, minimal insertion loss, high-frequency capabilities, mechanical stability and low phase noise.

\subsubsection{Model 4846-UU Cable} The 4846-UU cable (SMA Plug to SMA Plug, RG178B/U) was used for measuring and supplying any AC signal such as measuring LNA output noise. Its shielding and low-noise characteristics are crucial for test accuracy in noise-sensitive conditions.







\section{methods}

Characterization of the design involved conducting simulations at the post-layout stage, which included extracting parasitic capacitance and resistance. In this section, the details of these simulations will first be presented. Subsequently, the results obtained from benchtop testing of the fabricated LNA will be reported and discussed in the following sections.

\subsection{Frequency Bandwidth and Mid-Band Gain}\label{frequency-bandwidth}

Simulating the output amplitude in response to a 1V input AC signal (\emph{V\textsubscript{in}}), the AC simulator generates a frequency-dependent curve that resembles the amplitude characteristics depicted in the output Bode diagram shown in Fig. \ref{LNA_freq}.

The maximum amplitude of this curve is the mid-band differential gain (\emph{A\textsubscript{D}}), and the frequency range between -3~dB high-pass (\emph{f\textsubscript{HP-3dB}}) and lowpass cut-off frequencies (\emph{f\textsubscript{LP-3dB}}) is the bandwidth (BW) of the LNA.

\begin{figure}[]
  \includegraphics[width=0.4\textwidth,height=0.4\textheight,keepaspectratio]{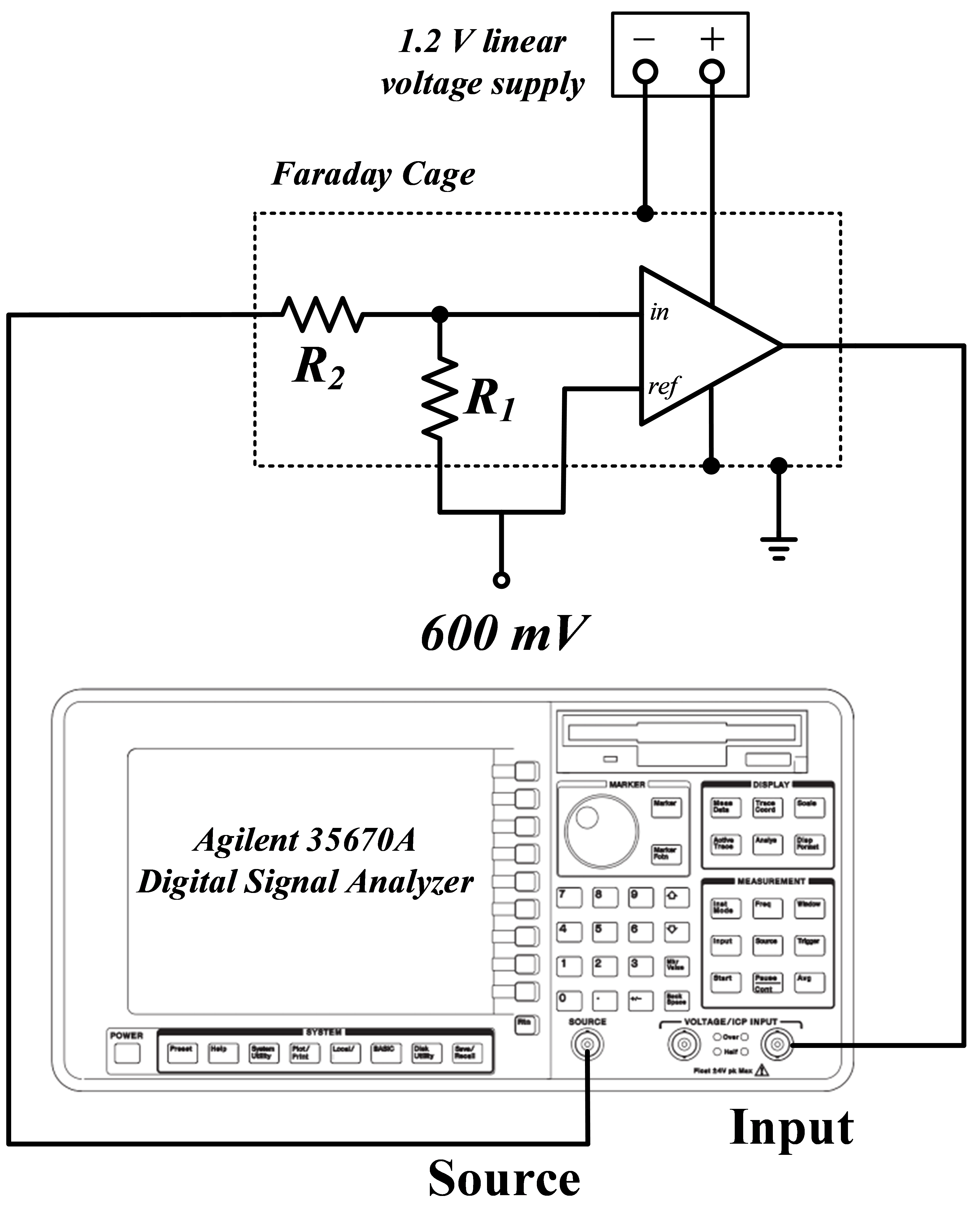}
  \centering
  \caption{Experimental schematic for measuring AC gain of the LNA}
    \label{gain}
\end{figure}

In terms of experimental measurements of the BW and the mid-band gain, Fig. \ref{gain} depicts the testbench used to measure these AC parameters. An Agilent 3670A, with its high measurement resolution and its input level noise less than -130 ~\si{\dB\volt_{rms}}/\si{\sqrt{Hz}} that would provide precise analysis of low-level biopotential signals, is utilized to characterize the LNA. In order to obtain the AC gain curve within the desired bandwidth, a frequency sweep is conducted using the [Sweep Sine] mode. 

Utilizing resistors (\emph{R\textsubscript{1}, R\textsubscript{2}}) to create an attenuation factor offers several advantages. Firstly, since most signal generators are incapable of generating signals below approximately \textasciitilde100 mV, an additional attenuation network becomes necessary to produce microvolt-level signals for testing purposes. Secondly, while the Agilent 35670a is capable of producing signals with amplitudes below a millivolt, this instrument can introduce noise that contributes to the overall output noise of the LNA. To mitigate this, an attenuation block can be added before the LNA, effectively reducing the noise from the AC source. Consequently, the input AC signal from the signal generator can be increased in amplitude while still achieving a noise reduction.

However, these resistors introduce another source of thermal noise: 

\begin{equation} \label{eq_thermal}
V_n^2 = 4kT \left(R_1 \parallel R_2\right) \cdot BW,
\end{equation}

where BW is the noise bandwidth of the device under test. Keeping R\textsubscript{1} \textless{100 \(\Omega\)} is typically sufficient to reduce this noise source to insignificance. Given that, an attenuation factor of 100 is required with a source voltage of 100 mV, R\textsubscript{2} = {10 k\(\Omega\)} would be chosen. To achieve a satisfactory resolution, the AC analysis can be performed by adjusting over 10 sampling points per frequency decade.

\subsection{Common Mode Rejection Ratio (CMRR)}

We stick with the traditional way of defining CMRR for op-amps, which goes like this:

\begin{equation} \label{eq_CMRR}
CMRR(dB) = 20log\left( \frac{A_{D}}{\left| A_{CM} \right|} \right),
\end{equation}

Here, \emph{A\textsubscript{D}} represents the differential gain, while \emph{A\textsubscript{CM}} stands for the common-mode gain of the amplifier.

We employ AC analysis in Cadence to measure \emph{A\textsubscript{CM}}, sticking within the amplifier's bandwidth. In simple terms, \emph{A\textsubscript{CM}} quantifies the size of the AC signal at the output when both differential inputs receive \emph{V\textsubscript{CM}} as their input AC signal, and there's no differential input signal. Then, we sweep through the signal frequency range from \emph{f\textsubscript{HP-3dB}} to \emph{f\textsubscript{LP-3dB}}. Ultimately, the lowest value on the right side of Eq.~\ref{eq_CMRR} represents the CMRR of the amplifier being studied.

\begin{figure}[]
  \includegraphics[width=0.4\textwidth,height=0.4\textheight,keepaspectratio]{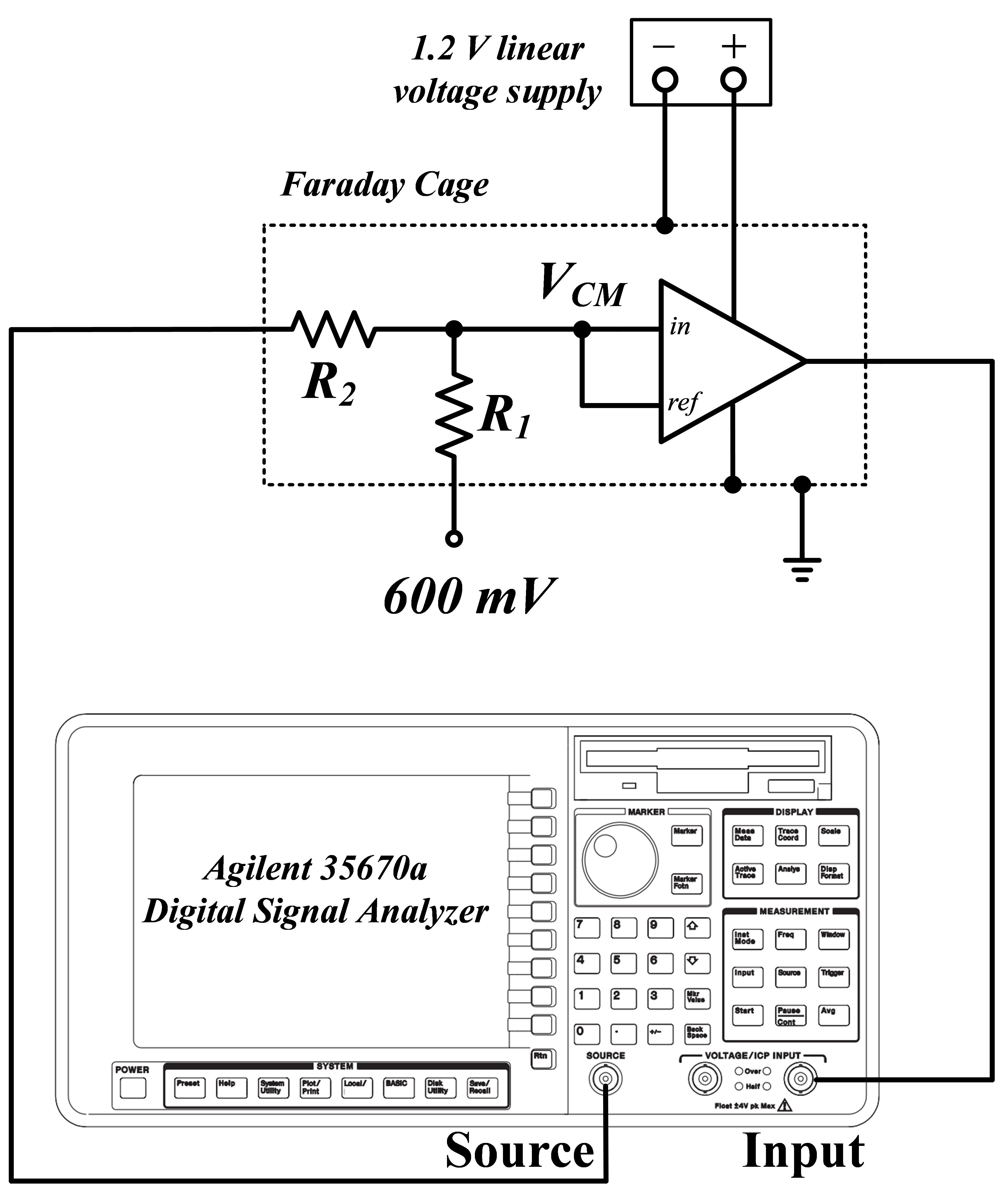}
  \centering
  \caption{Testbench for measuring CMRR.}
    \label{CMRR}
\end{figure}

To practically measure \emph{A\textsubscript{CM}}, the same procedure as measuring mid-band gain was followed, but input terminals of the LNA (\emph{V\textsubscript{in}} and \emph{V\textsubscript{ref}}) were both tied to have the same input signal \emph{V\textsubscript{CM}}~\cite{gosselin2011circuits} (Fig. \ref{CMRR}). This is used to generate a curve of \emph{A\textsubscript{CM}}'s amplitude versus frequency over frequency from \emph{f\textsubscript{HP-3dB}} to \emph{f\textsubscript{LP-3dB}} in order to find the maximum value.

\subsection{Power Supply Rejection Ratio (PSRR)}

We use Cadence Spectre APS' AC analysis tool to explore PSRR. This involves sweeping the signal frequency across the amplifier's bandwidth. To calculate PSRR, we use this formula:

\begin{equation} \label{eq_PSRR}
PSRR = 20\log\frac{A_{v}}{A_{dd}\left( V_{in} = 0 \right)}
\end{equation}

In simple terms, this equation compares the differential gain (\emph{A\textsubscript{v}}) to the gain influenced by power-supply ripple when the differential input is at zero (\emph{A\textsubscript{dd}}) ~\cite{allen2011cmos}.

To introduce power-supply ripple, we insert an AC source separately to the power supply rails +\emph{V\textsubscript{dd}} and \emph{-V\textsubscript{SS}}. This action enables the calculation of PSRR+ and PSRR- without applying an AC signal to the amplifier's differential input.

\begin{figure}[]
  \includegraphics[width=0.4\textwidth,height=0.4\textheight,keepaspectratio]{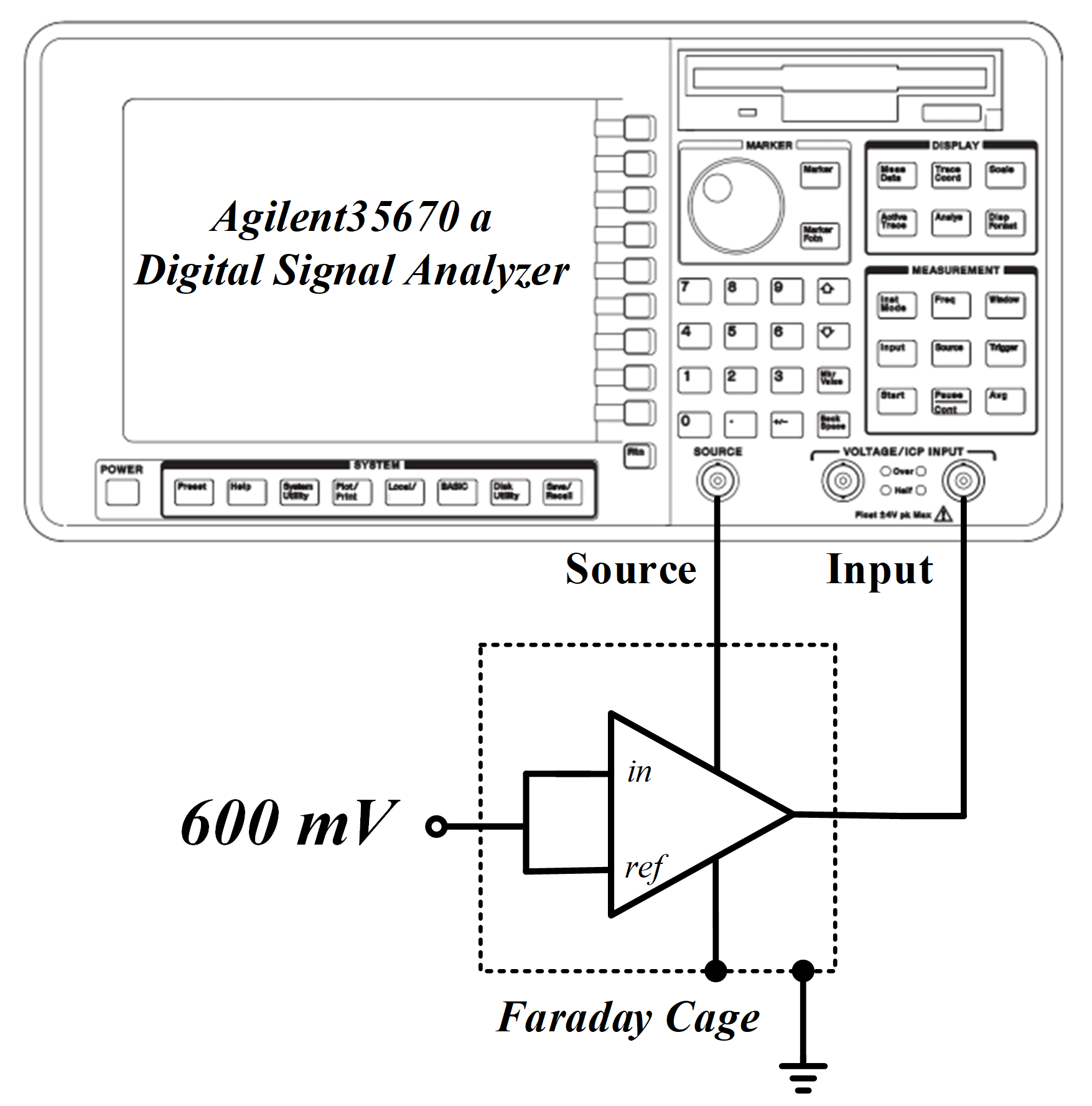}
  \centering
  \caption{The diagram of the test set-up to obtain the gain of the LNA when the AC input is from the supply (\( A_{dd}\)). }
    \label{Sch_PSRR}
\end{figure}

Fig. \ref{Sch_PSRR} is one of the possible experimental testbench for measuring the PSRR. It is necessary to verify the DC and AC levels of the power signal and signal generator delivered to the LNA. What is expected is the curve of \emph{A\textsubscript{dd}}'s amplitude per frequency, including bandwidth from \emph{f\textsubscript{HP-3dB}} to \emph{f\textsubscript{LP-3dB}}, to find the maximum of \emph{A\textsubscript{dd}}.

\subsection{Input-Referred Noise}

Simulation of input-referred noise in V\textsubscript{rms} is achievable through Cadence Virtuoso's noise analysis tools (Noise Summary). This process involves probing the LNA's input and output nodes under a specified DC bias in order to extract the RMS noise within the system's bandwidth. To encompass the complete RMS noise – incorporating both thermal and flicker noise – the RMS output noise must be divided by the LNA's bandpass gain.

\begin{figure}[]
  \includegraphics[width=0.45\textwidth,height=0.45\textheight,keepaspectratio]{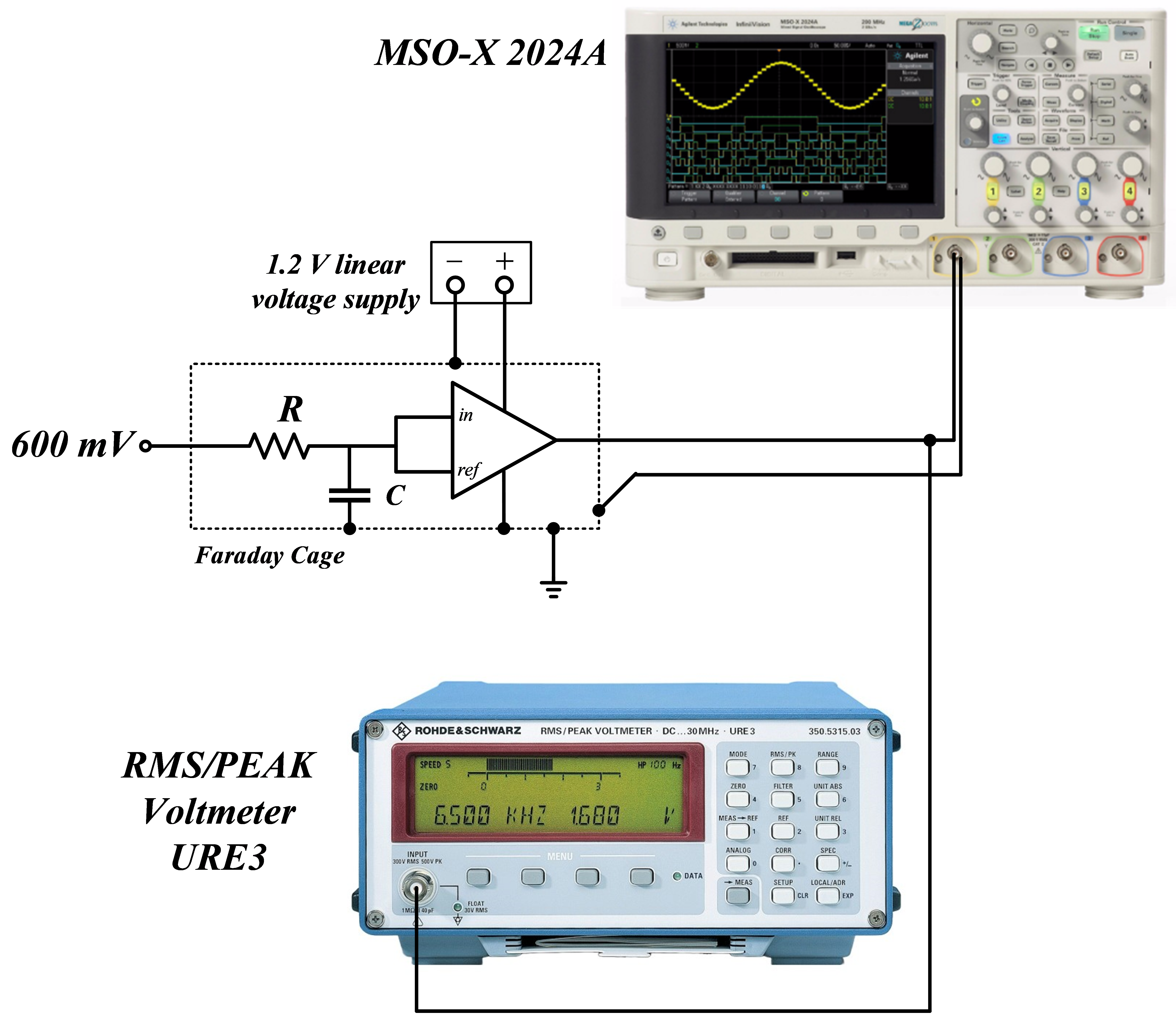}
  \centering
  \caption{Experimental setup with input low-pass filter to measure the output noise of the fabricated LNA.}
    \label{Noise_measure}
\end{figure}

The input-referred noise has been measured as depicted in Fig. \ref{Noise_measure}. An RC low-pass filter is used to cancel any noise from bias circuits with R=10 \( k\Omega \), C=10 \( \mu F\). Then, to have a precise noise measurement of the fabricated LNA, in terms of V\textsubscript{rms}, both a digital and an analog measurement instrument are applied in parallel.

The Agilent MSO-X 2024A, featuring mathematical functions and real-time calculation capabilities, is utilized to capture the Fast Fourier Transform (FFT) of the LNA when only the DC bias is present on the inputs. The oscilloscope provides the RMS voltage for each FFT sample. Then. the FFT curve should be converted to a Power Spectrum Density (PSD) according to Eq. \ref{PSD}.

\begin{equation} \label{PSD}
PSD (X_i) = \frac{{(X_i)^2}}{{\Delta f}},
\end{equation}
where \(X\textsubscript{i}\) is an FFT sample in RMS scale, and \(\Delta f\) is the sampling frequency resolution.

The RMS value of the input-referred noise is the integrated area under this curve and is divided by mid-band gain over the desired bandwidth. Numerical integration can be performed by simply summing the FFT samples between frequencies \emph{f\textsubscript{HP-3dB}} and \emph{f\textsubscript{LP-3dB}}, inclusively, according to Eq.~\ref{VNRMS}.

\begin{equation} \label{VNRMS}
V_{\text{in,rms}} = \frac{{\sum_{i=f_{\text{HP-3dB}}}^{i=f_{\text{LP-3dB}}} \text{PSD}(X_i)}}{{A_v}}
\end{equation}

RMS voltmeter URE3 is an analog measurement instrument with less than 10~\si{\micro\volt_{rms}} input noise. Thus, this provides testing means equivalent to the digital approach (Fig. \ref{Noise_measure}). Choosing an AC-RMS mode RMS meter gives the RMS value of the output noise of the LNA in a selected BW. Then, the measured value should be divided by \emph{A\textsubscript{v},} to obtain the RMS input reference noise of the LNA.

\subsection{DC Offset Cancellation}

In assessing the DC offset cancellation performance of the proposed LNA, the methodology involves measuring the maximum DC variation range of the output. This measurement is conducted with a variable input offset applied to the LNA input, ranging from GND to Vdd. The observed DC variation of the input at very low frequencies is referred to as dynamic offset, providing insights into the efficacy of the DC offset cancellation mechanism.

\subsection{Linearity}

Numerous articles rely on total harmonic distortion (THD) to explain linearity. However, in this study, our central focus for spike-recording applications is the reduction in gain caused by interference like electromagnetic interference or low-frequency local field potentials, leading to varying gains over time~\cite{zhang2012design}.

Hence, we propose that assessing the -1 dB gain compression point (roughly 89\% of voltage gain) is more practical for characterizing these amplifiers, compared to using THD.

 The test set-up for measuring the -1 dB gain compression is the same as that for measuring the mid-band gain (Fig. \ref{gain}). The LNA should be connected to a signal generator and a spectrum analyzer or power meter. Then the signal generator is set to output a sine wave signal at a representative frequency. By increasing gradually the input signal voltage, we identify the point where the gain drops by 1 dB from the linear gain~\cite{tilden1999overview}. This is the -1 dB gain compression point. The linearity performance can be evaluated by comparing the input voltage for dB gain compression.

\subsection{Efficiency}

For benchmarking our amplifier's noise and power capabilities against others, we employ the noise efficiency factor (NEF)~\cite{steyaert1987micropower}.

\begin{equation} \label{NEF}
\textrm{NEF} = v_{in,rms} \sqrt{\frac{2I_{tot}}{\pi \cdot U_t  \cdot  4kT  \cdot  BW}},
\end{equation}
where \emph{I\textsubscript{tot}} is the total amplifier supply current, and it is measurable by a Model 6487 Picoammeter/Voltage Source. This instrument has a digit resolution of 10 fA. \emph{U\textsubscript{t}} is the thermal voltage, BW is the amplifier bandwidth, and \emph{V\textsubscript{in,rms}} is the amplifier's input-referred RMS voltage noise and it is integrated up to \emph{f\textsubscript{LP-3dB}}. \emph{k}, the Boltzmann constant is defined as $1.38\times 10^{-23}$~J$\cdot$K$^{-1}$. The Figure of Merit (FOM) scales the input-referred RMS noise voltage to match that of an ideal single-transistor bipolar amplifier with equivalent current consumption and bandwidth.

When comparing two circuits operating at the same supply voltage, NEF serves as a suitable measure to assess the power-to-noise balance. However, if two amplifiers possess identical total current and noise yet operate at different VDD values, their NEF might be the same, even though their power usage differs. This indicates that NEF alone isn't adequate for evaluating power efficiency. To address this concern, a more direct evaluation of overall power consumption can be achieved using the Power Efficiency Factor (PEF) metric ~\cite{muller20110}. PEF normalizes the product of noise power and total power, resulting in \(PEF=NEF\textsuperscript{2}\cdot V_{dd}\).

Although the PEF is adequate to compare the performance of the amplifiers in terms of power consumption and with the same die area, it does assess the merits of the amplifiers when silicon occupation surface is an essential parameter. In this work, the priority is designing a very small-sized LNA to have a very dense neural recorder. Therefore, we propose a different metric designated Area Efficiency Factor (AEF) and expressed as
\begin{equation}
    \textrm{AEF}=\textrm{PEF} \times \frac{A}{10^{-6}},
\end{equation}
where $A$ is the total silicon area of the LNA including its capacitors.

In the following section, we present an exhaustive compilation of findings obtained through diverse methodologies employed for the measurement of LNA parameters.

\section{RESULTS}

The biopotential amplifier is fabricated in a 28 nm process from TSMC Semiconductor Manufacturing Company. We use Metal-Insulator-Metal (MIM) capacitors for their high density, good linearity, and low substrate capacitance. The circuit operates from a 1.2 V supply voltage.

\begin{figure}[]
  \includegraphics[width=0.45\textwidth,height=0.45\textheight,keepaspectratio]{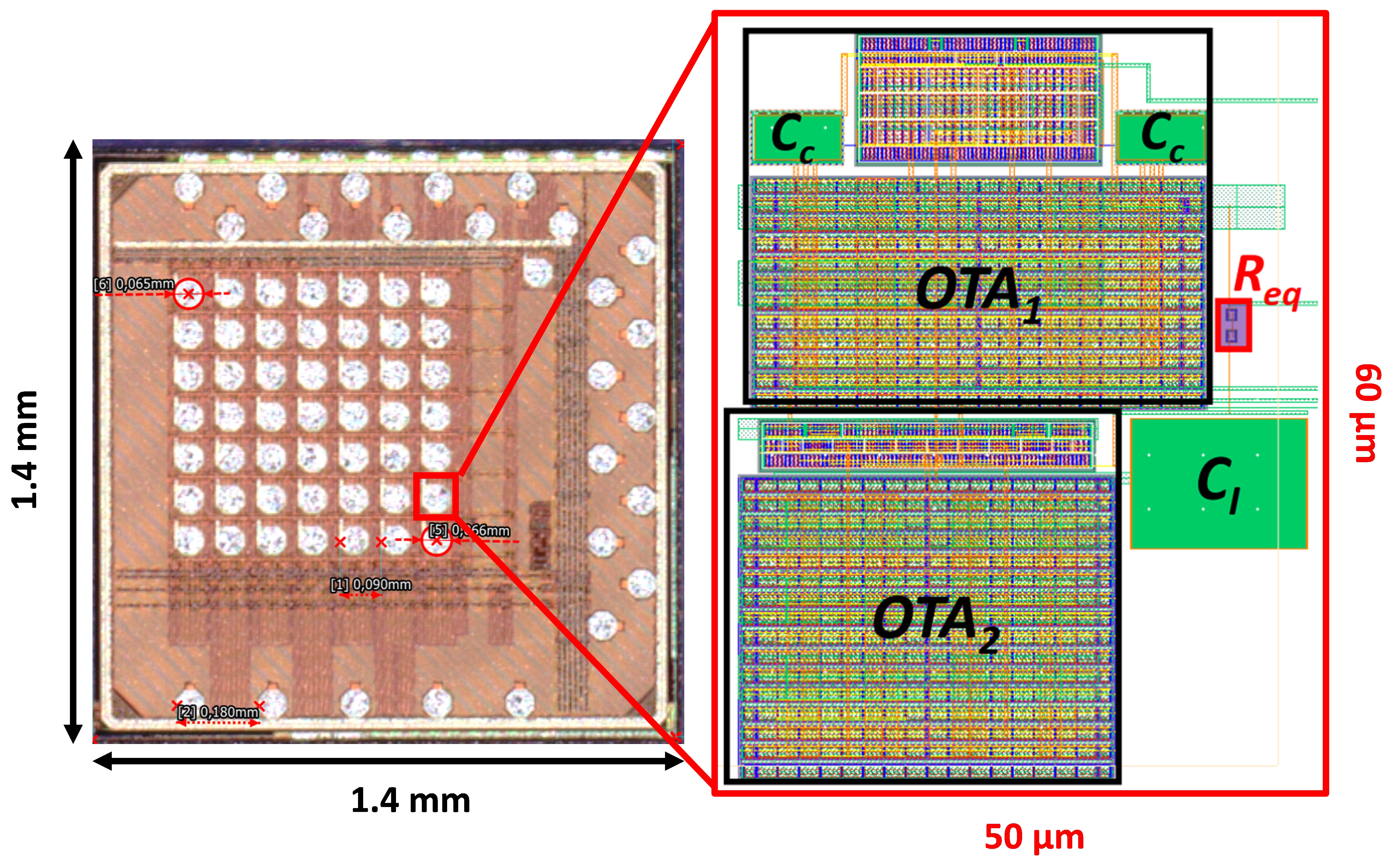}
  \centering
  \caption{ Micrograph of the ASIC featuring 49 recording channels on the left, alongside the layout of the proposed LNA designed in 28 nm CMOS technology on the right.}
    \label{Migrograph}
\end{figure}

The microphotograph of the fabrication chip is shown in Fig. \ref{Migrograph} and the LNA occupies an area of 2500~\si{\micro\meter\squared}.

\subsection{Frequency Bandwidth and Mid-Band Gain}

\begin{figure}[]
  \includegraphics[width=0.45\textwidth,height=0.45\textheight,keepaspectratio]{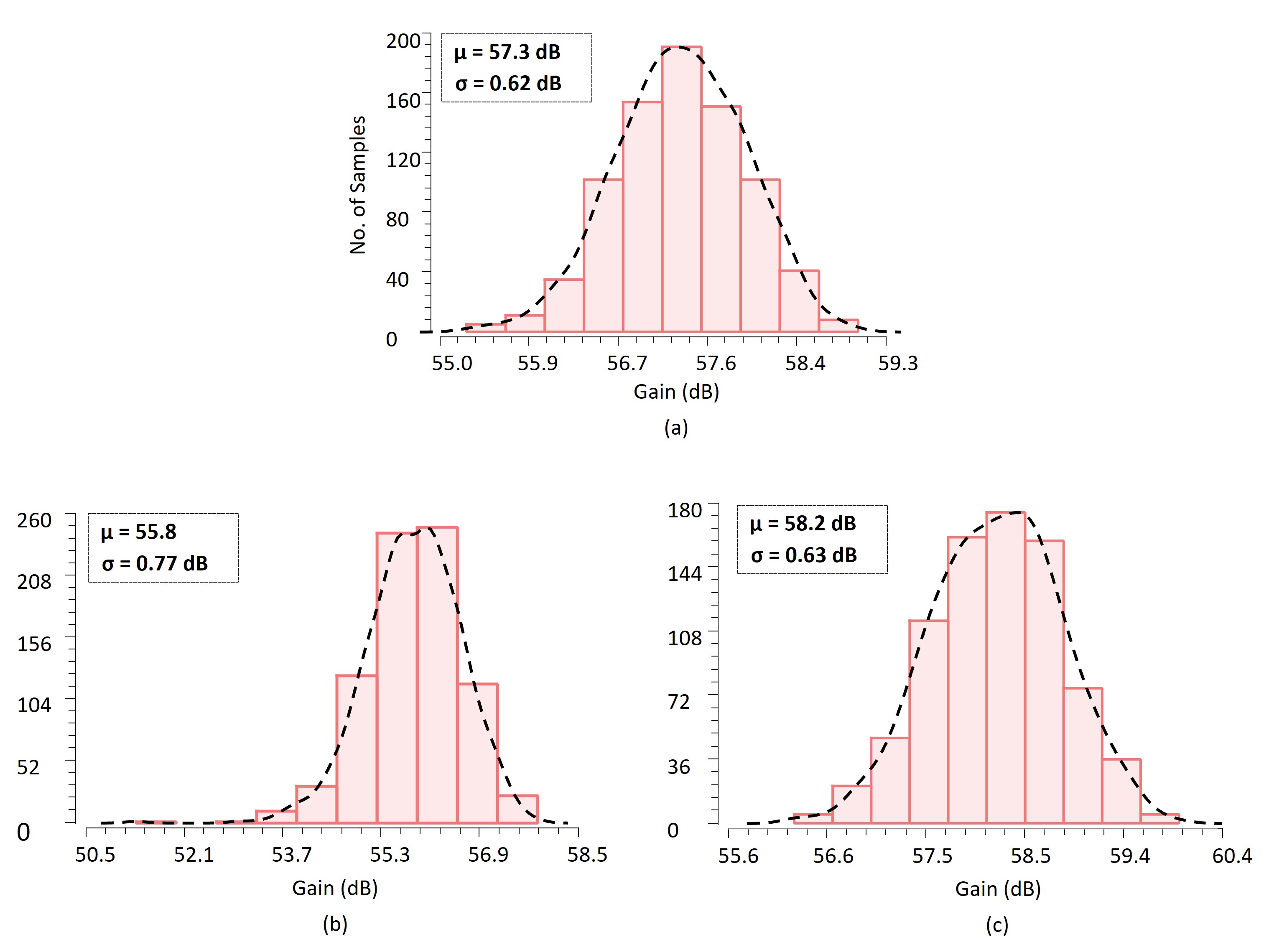}
  \centering
  \caption{Statistical distribution of the mid-band gain of LNA. (a) \emph{V\textsubscript{DD}} = 1.2 V, (b) \emph{V\textsubscript{DD}} = 1.1 V, (c) \emph{V\textsubscript{DD}} = 1.3 V}
    \label{montcarlo}
\end{figure}

\begin{figure}[]
  \includegraphics[width=0.4\textwidth,height=0.4\textheight,keepaspectratio]{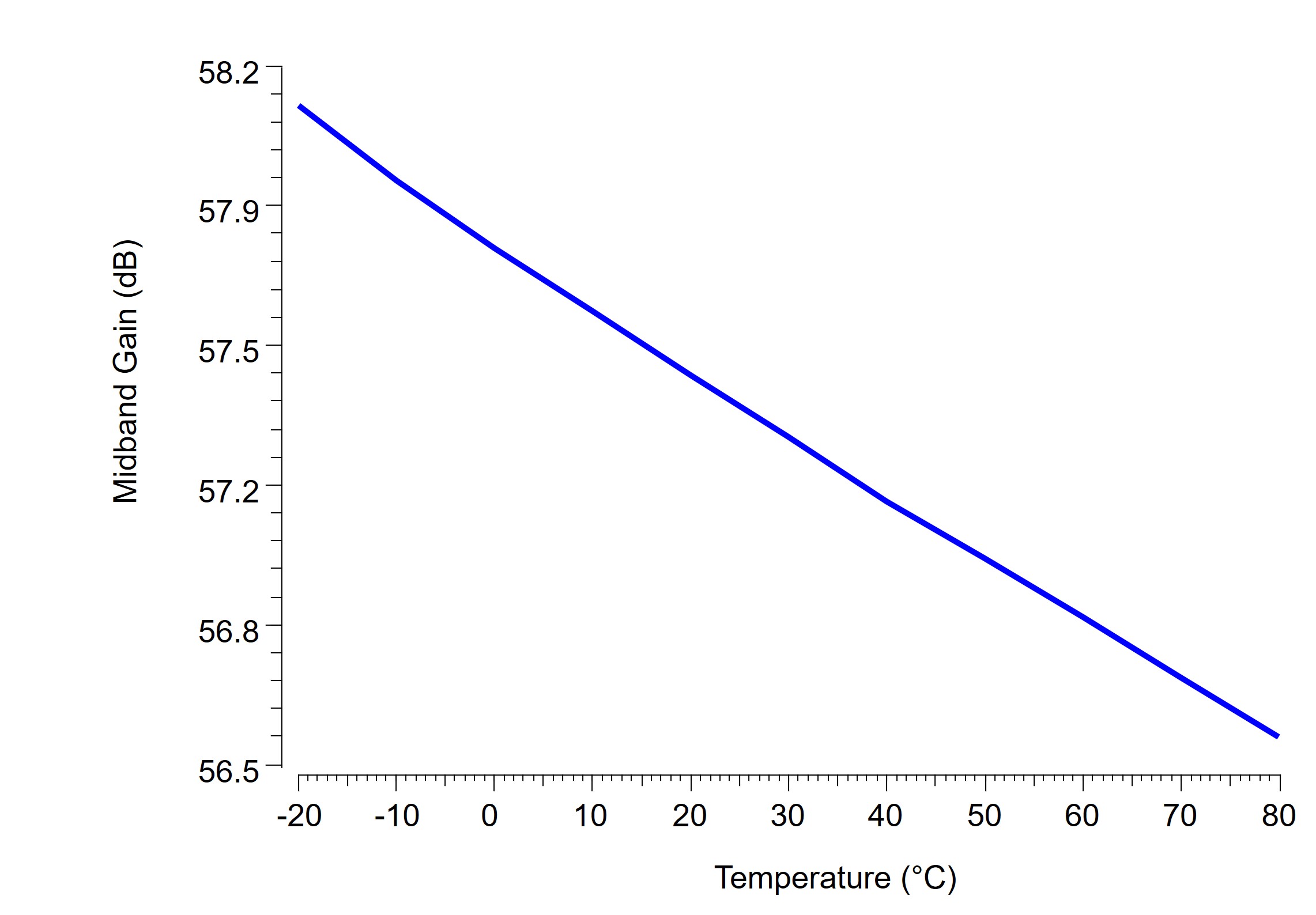}
  \centering
  \caption{Mid-band gain of the LNA depending on temperature.}
    \label{temp}
\end{figure}

The simulation resulted in \emph{f\textsubscript{HP-3dB}} = 600 Hz and \emph{f\textsubscript{LP-3dB}} = 7 kHz for the Typical-Typical (TT) process corner. Fig. \ref{montcarlo} shows the statistical distribution of mid-band gain obtained using 800 Monte Carlo simulations. The results include the local and global mismatches depending on process corners. When VDD changes by 10\%, the average gain varies from 55.8 to 58.2~dB. Moreover, the LNA mid-band gain depending on temperature is investigated using circuit simulation (Fig. \ref{temp}).

\begin{figure}[]
  \includegraphics[width=0.4\textwidth,height=0.4\textheight,keepaspectratio]{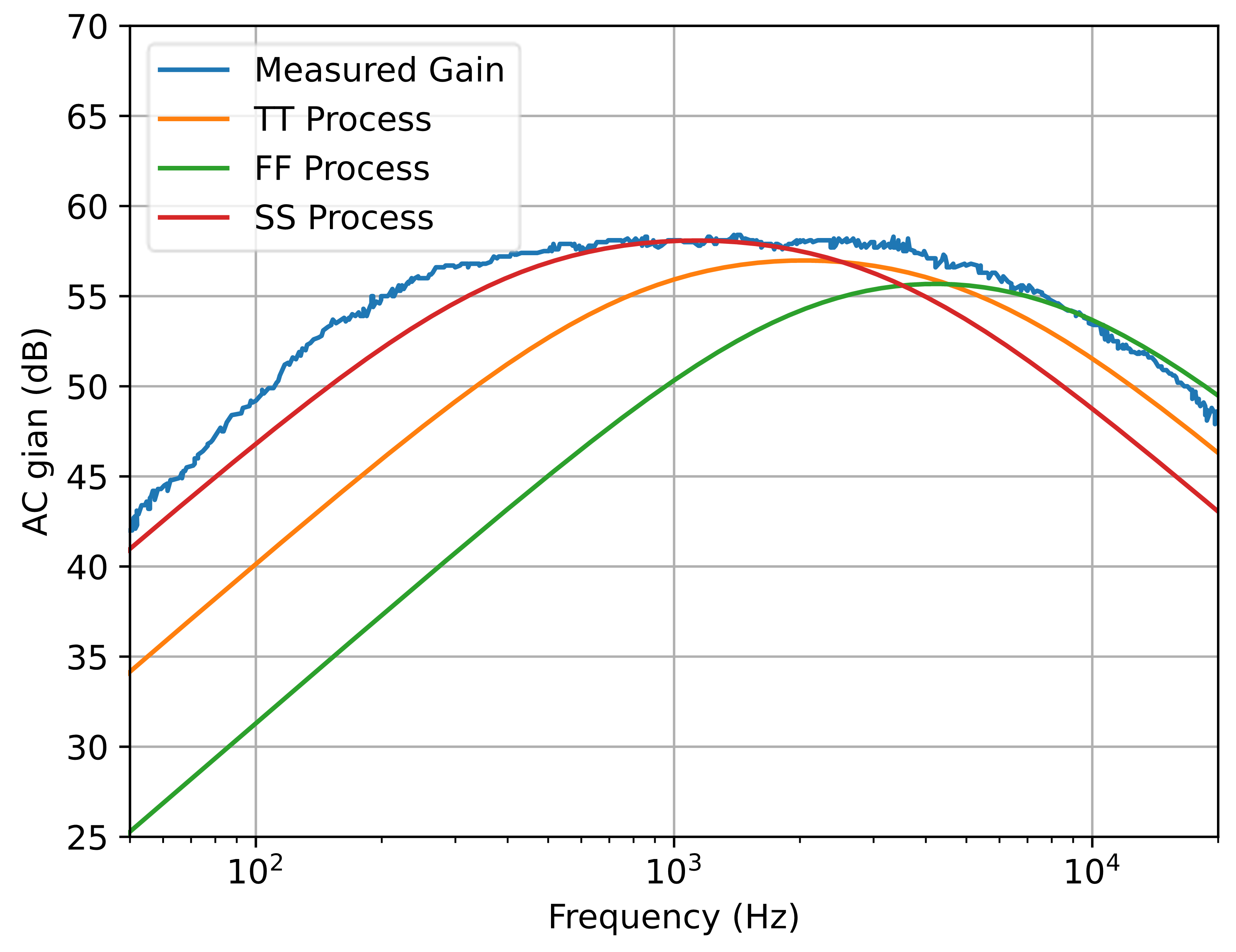}
  \centering
  \caption{Comparison of Measured and Simulated Gain Magnitude Responses of the LNA in TT, FF, and SS Process.}
    \label{Gain_Curve}
\end{figure}

According to the measured AC response (Fig. \ref{Gain_Curve}), the mid-band gain is 58 dB. The -3 dB high-pass corners occur at approximately 150 Hz, and the -3~dB low-pass corner is 7.1 kHz. The gain variation measured across 5 LNAs on different chips was less than 1 dB.

The difference between the simulated \emph{f\textsubscript{HP-3dB}} and the measured value is evident. As shown in Fig. \ref{LNA_freq}, this can be attributed to the strong influence of variations in \(R_{eq}\), as well as the gain of OTA\textsubscript{1} and OTA\textsubscript{2}.

\subsection{CMRR}

Single-ended LNAs with active feedback always suffer from poor PSRR and CMRR. CMRR for all the amplifiers is measured via post-layout simulation. A value of 78 dB was simulated for \emph{CMRR\textsubscript{OTA1}} and 70 dB for \emph{CMRR\textsubscript{OTA2}}. However, the CMRR of the LNA is 22 dB. To suppress the noise originating from the integrated reference voltage source, the fabricated LNA incorporates a low-pass RC filter within the ASIC. Due to the inclusion of this filter, it is not feasible to measure the CMRR of our ASIC without disturbing its functionality.

\subsection{PSRR}
Here are the outcomes from the PSRR simulation.

$A_v=55$ dB, $A_{dd+} = 16$ dB,
$A_{dd-} = 10$ dB,

PSRR = min \{PSRR\textsubscript{-} = 45 dB, PSRR\textsubscript{+}
= 39 dB\} = 39 dB.

\begin{figure}[]
  \includegraphics[width=0.4\textwidth,height=0.4\textheight,keepaspectratio]{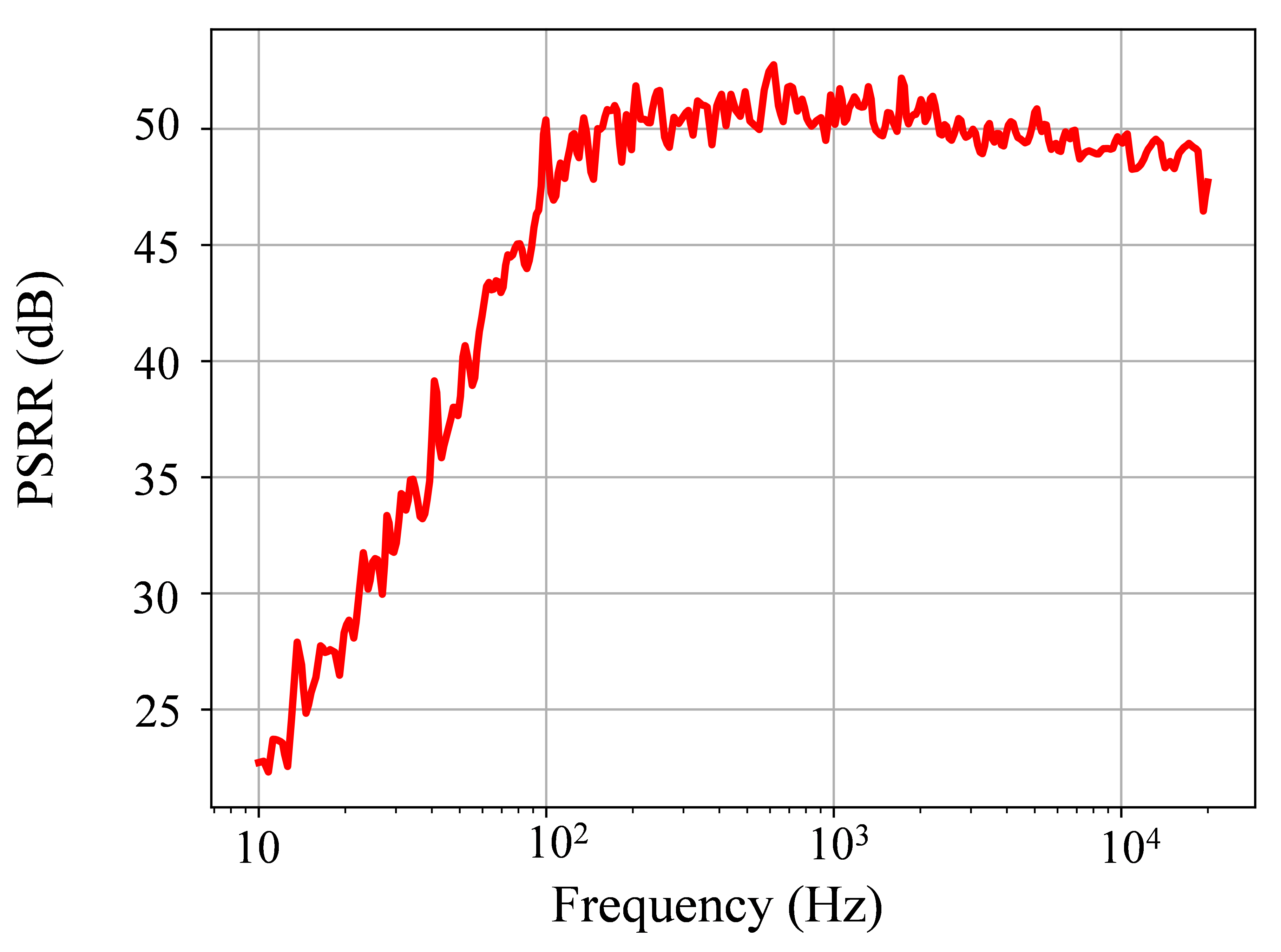}
  \centering
  \caption{Measured PSRR of the LNA. }
    \label{PSRR_curve}
\end{figure}

 Fig. \ref{PSRR_curve} shows the measured PSRR of the fabricated LNA. According to this curve and Eq.~\ref{eq_PSRR}, the worst measured PSRR of the LNA in the bandwidth (150 Hz to 7.1 kHz) is 50 dB.

\subsection{Noise Analysis}

In this work, the input-referred noise from the post-layout simulation is 8.4 µV\textsubscript{rms} for TT process corner when the mid-band gain is 57 dB for a 600 Hz to 7 kHz bandwidth.

\begin{figure}[]
  \includegraphics[width=0.45\textwidth,height=0.45\textheight,keepaspectratio]{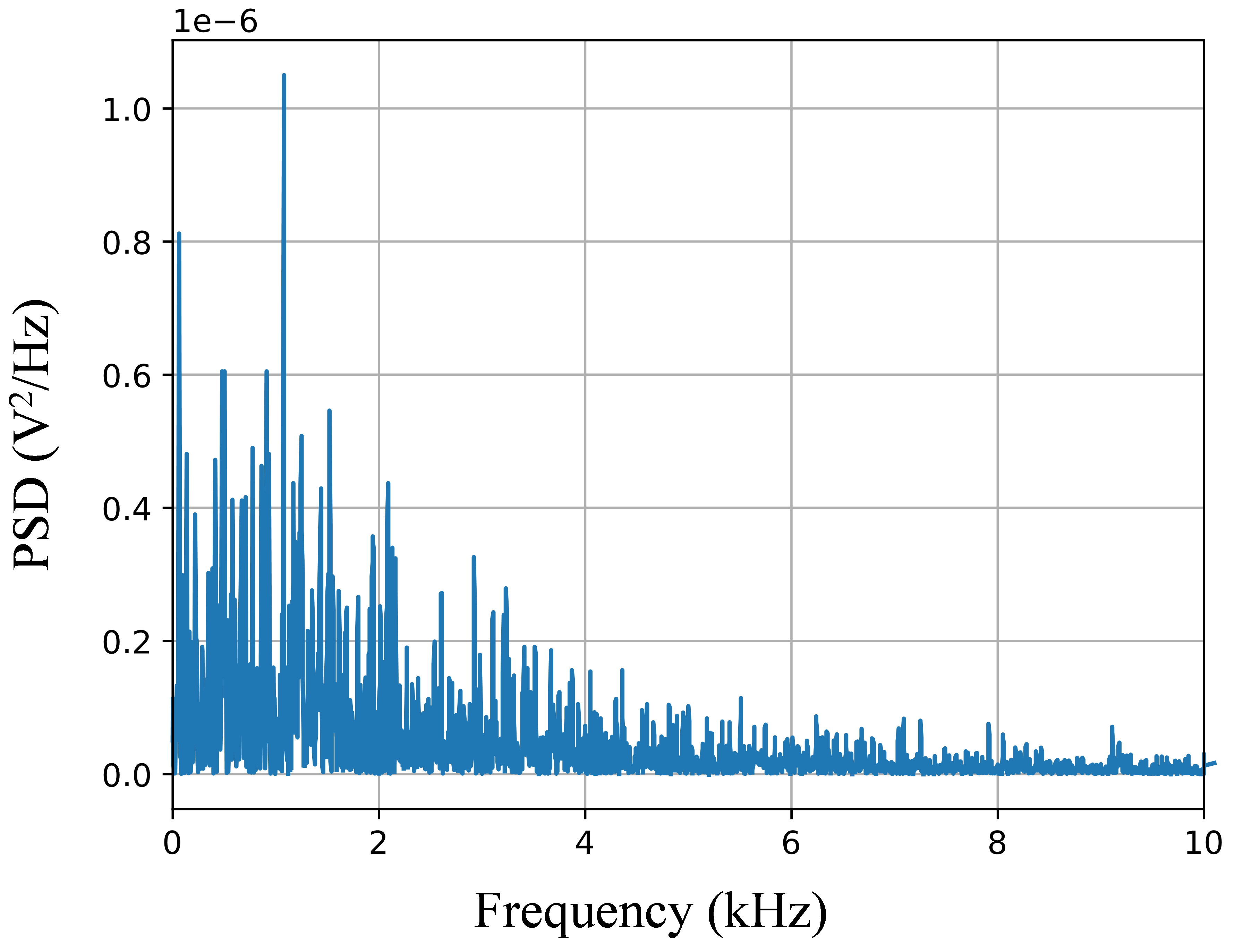}
  \centering
  \caption{PSD of measured FFT output noise.}
    \label{PSD_noise}
\end{figure}

In Fig. \ref{PSD_noise}, the Power Spectrum Density (PSD) of the noise in the LNA output is depicted, which is calculated from the FFT analysis of the measured noise. The RMS value of the output and input noise in the bandwidth is 13.4 mV\textsubscript{rms} and 15.8 µV\textsubscript{rms}, respectively, and the frequency of the corner is almost 5 kHz. Here, the output noise of the instrument is 1 mV\textsubscript{rms}, which is not significant compared to the output noise of the LNA. 

The noticeable distinction between the noise analysis results derived from post-layout simulation and the measured noise can be attributed to the additional noise induced by the integrated voltage source bias applied to the input of the LNA. Regrettably, complete elimination of this noise is not possible in our measurements. Furthermore, due to the lower \emph{f\textsubscript{HP-3dB}}, the bandwidth encompasses a greater amount of flicker noise in comparison to the simulated noise value.

\begin{table*}[!ht]
\centering
\caption{Comparison table for different process corner simulations.}
\label{Corners}
\resizebox{\textwidth}{!}{%
\begin{tabular}{|l|c|c|c|c|c|}
\hline
\textbf{Process Corner} & \textbf{TT}    & \textbf{FF}         & \textbf{FS}      & \textbf{SF}       & \textbf{SS}      \\ \hline
\textbf{Input-referred noise in BW (µVrms)} & 8.4  & 8.5  & 8.8  & 8    & 8.5  \\ \hline
\textbf{Midband gain (dB)}                  & 57   & 55.7 & 56.7 & 57.1 & 58.1 \\ \hline
\textbf{Output noise in BW (mVrms)}         & 6    & 5.2  & 5.9  & 5.8  & 6.9  \\ \hline
\textbf{Power consumption (µW)}             & 3.55 & 3.71 & 2.36 & 3.61 & 2.34 \\ \hline
\textbf{BW}             & 600 Hz – 7 kHz & 1.45 kHz - 12.2 kHz & 370 Hz - 5.2 kHz & 1.1 kHz - 8.4 kHz & 330 Hz - 3.9 kHz \\ \hline
\end{tabular}%
}
\end{table*}

\subsection{DC Offset Cancellation}

\begin{figure}[]
  \includegraphics[width=0.45\textwidth,height=0.45\textheight,keepaspectratio]{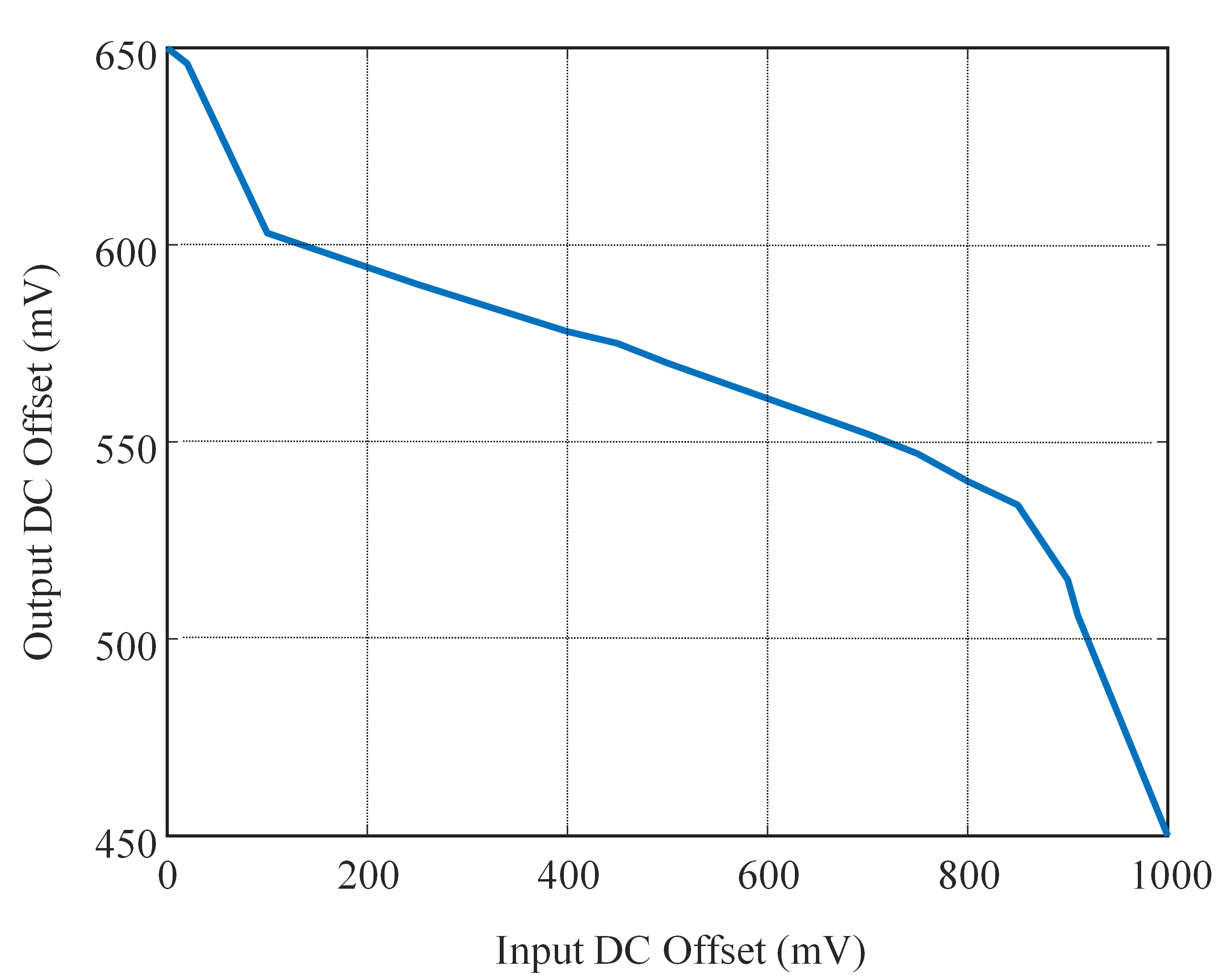}
  \centering
  \caption{Measured dynamic DC offset and DC cancellation performance of the LNA.}
    \label{DC_Cancellation}
\end{figure}

In Fig.~\ref{DC_Cancellation}, the measured output dynamic DC offset is illustrated concerning DC input voltage bias ranging from 0 to 1.2 volts. The graph indicates that the LNA avoids saturation even with a variation in DC input of approximately 1 volt, spanning from 0 to 910~mV. As depicted in the diagram in Fig.~\ref{DC_Cancellation}, the DC cancellation gain of the LNA is determined to be -22~dB.

\subsection{Linearity}

The measured result indicates that the -1 dB gain compression point is observed at an input level of 1.4 mV.

\subsection{Efficiency}

To assess the LNA current, a dedicated supply pin is utilized, providing power to an LNA test alongside its comparator. Initially, the current of this LNA-comparator combination was measured for 5 ASICs, yielding an average of 4.9 µA. Subsequently, through simulation of the comparator in all 5 process corners, the average comparator consumption was determined to be 2.1 µA. By deducting the comparator's current from the overall measured current, the LNA's current consumption can be calculated at 2.8 µA when supplied with 1.2 V.

Based on the measured BW of 7 kHz and the measured input-referred noise level of 15.8 µV\textsubscript{rms}, the NEF of the LNA was determined to be 10.6. With a supply voltage of 1.2 V, the PEF was calculated to be 134.8. Then, by assigning a layout area of 2500~\si{\micro\meter\squared} to the LNA, an Area Efficiency Factor (AEF) of 0.34 was obtained.

Besides the TT process corner, four other process corners, namely Fast-Fast (FF), Fast-Slow (FS), Slow-Fast (SF), and Slow-Slow (SS) are reported in Table \ref{Corners} as simulations results.



\section{Exploration and Comparative Examination}

\subsection{Neural Signal Recording Using 28 nm CMOS Technology}

Designing for low noise using very short-channel technologies is quite challenging. This is mainly because the way transconductance behaves in the 28 nm CMOS technology is not straightforward – it doesn't just increase with more drain current or transistor area (\emph{WL}), which goes against Eq.~\ref{eq_Noise_Fl&ther}. Additionally, the suggested architecture, where a feedback loop is directly connected, has trouble with not-so-great PSRR and CMRR.

Making things even more complex, the noise from OTA\textsubscript{2}'s output becomes the input for OTA\textsubscript{1}. As a result, this noise keeps getting amplified by the LNA.

Furthermore, due to the high impedance of the electrode on one input of the feedforward amplifier (OTA\textsubscript{2}), the LNA exhibits inadequate CMRR. To enhance both PSRR and CMRR, a fully-differential architecture can be employed for the proposed LNA. However, this approach results in a twofold increase in power consumption, die area, and input-referred noise of the LNA.

\subsection{Comparative Assessment with Previous Studies}

Table \ref{Tab_Compare} presents a comparison between this study and previous works on DC-coupled LNAs that utilize technologies featuring longer channel lengths. While the adoption of a single-ended architecture may seem unconventional, it is a deliberate choice driven by considerations of practicality and efficiency. Single-ended LNAs exhibit greater sensitivity to common noises, such as CMRR and PSRR. However, in the context of our research, a fully differential architecture would lead to a bulky implementation. This choice becomes a critical consideration when emphasizing the urgent necessity for an ultra-compact design accommodating 49 channels within a limited 1 mm² space. The outcomes of our investigation demonstrate a remarkable reduction in chip size by a factor of at least 5, accompanied by a substantial decrease in energy consumption, as compared to the circuits detailed in Table \ref{Tab_Compare}.

It is crucial to note that, despite the reduction in chip size and energy consumption, there is a slight increase in noise level. The table further provides well-estimated NEF and PEF metrics, along with the area efficiency factor. These additional insights contribute to a more thorough understanding of the trade-offs made in our design choice.

\begin{table*}[]

\centering
\caption{COMPARISON TABLE WITH PRIOR ARTS.}
\label{Tab_Compare}
\begin{tabular}{|l|c|c|c|c|c|c|}
\hline
\textbf{Reference} &
  \multicolumn{1}{l|}{\textbf{This work}} &
  \multicolumn{1}{l|}{\textbf{VLSI'23\cite{huang2023noise}*}} &
  \multicolumn{1}{l|}{\textbf{TCAS-II'21}\cite{pham20200}} &
  \multicolumn{1}{l|}{\textbf{TBioCAS'07\cite{gosselin2007low}}} &
  \multicolumn{1}{l|}{\textbf{JSSC'11\cite{muller20110}}} &
  \multicolumn{1}{l|}{\textbf{TBioCAS'07\cite{wattanapanitch2007energy}}} \\ \hline
\textbf{Tech.}                           & 28 nm    & 180 nm & 180 nm & 180 nm   & 65 nm   & 180 nm    \\ \hline
\textbf{VDD (V)}                         & 1.2      & 1.8    & 0.8    & 1.8      & 0.5     & 1.8       \\ \hline
\textbf{Power (µW)}                      & 3.4      & 13.9   & 0.52   & 8.6      & 5.13    & 20        \\ \hline
\textbf{Area (mm\textsuperscript{2})}                      & 0.0025   & 0.085  & 0.24   & 0.05     & 0.013   & 0.03      \\ \hline
\textbf{Input Impedance~\si{\Omega}} &
  \begin{tabular}[c]{@{}c@{}}280 M\\ @1 kHz\end{tabular} &
  \begin{tabular}[c]{@{}c@{}}64 M\\ @60 Hz\end{tabular} &
  N/A &
  N/A &
  N/A &
  N/A \\ \hline
\textbf{Electrode Offset Tolerance (mV)} & 910      & 50     & N/A    & 900      & 100     & N/A       \\ \hline
\textbf{Gain (dB)}                       & 58       & 40     & 40     & 50       & -       & 49-66     \\ \hline
\textbf{BW (Hz)}                         & 150-7.1k & 1-100  & 800    & 100-9.1k & 300-10k & 350-11.7k \\ \hline
\textbf{Noise (µV\textsubscript{rms})}                   & 15.8     & 0.59   & 1.1    & 5.6      & 4.9     & 5.4       \\ \hline
\textbf{PSRR (dB)}                       & 48       & N/A    & 75     & 52       & 50      & 72        \\ \hline
\textbf{NEF}                             & 12       & 6.4    & 2.1    & 4.9      & 6       & 7         \\ \hline
\textbf{PEF}                             & 175      & 73.7   & 1.2    & 43.2     & 18      & 58.8      \\ \hline
\textbf{AEF}                             & 0.43     & 6.3    & 0.29   & 2.16     & 0.23    & 1.06      \\ \hline
\end{tabular}
\end{table*}

\section{conclusion}

In conclusion, this paper introduced a DC-coupled biopotential amplifier that replaces traditional DC-blocking capacitors with analog feedback. This feedback mechanism effectively senses and cancels the DC offset at the input by monitoring the output of the amplifier. The implementation of this approach, combined with the utilization of a 28 nm CMOS node, resulted in a significant reduction in the required area. As a result, a recording channel could be seamlessly integrated into a compact 100 µm x 100 µm pitch for a 49-channel neural recording ASIC. Moving forward, our future work will focus on the development of a comprehensive recording system on a single chip, aiming to achieve an even denser arrangement of channels and ultimately enhance the spatial resolution of the recorded neural data.


%


\section*{Acknowledgment}

Our heartfelt thanks to Gabriel Martin-Hardy for their exceptional wire-bonding expertise, and to Gabriel Gagnon-Turcotte for providing invaluable laboratory support during the experimental tests. Their technical assistance has been instrumental in the success of this project, and we are deeply grateful for their contributions.


\ifCLASSOPTIONcaptionsoff
  \newpage
\fi



%

\bibliographystyle{IEEEtran}
\bibliography{references}




%








\end{document}